\documentclass[prl,twocolumn,nofootinbib,amsmath,amsfonts,amssymb,showpacs]{revtex4}

\usepackage[usenames]{color}
\usepackage{epsfig,graphicx,bm,color}

\usepackage{natbib}
\newcommand{\be}{\begin{equation}}
\newcommand{\ee}{\end{equation}}
\newcommand{\bea}{\begin{eqnarray}}
\newcommand{\eea}{\end{eqnarray}}

\newcommand{\sigv}{\mbox{$\langle \sigma v \rangle $}}

\begin{document}

\title{Intermediate Mass Black Holes and Nearby Dark Matter Point Sources:\\ 
  A Critical Reassessment}

\author{Torsten Bringmann}
\affiliation{OKC, % Oskar Klein Centre for Cosmo Particle Physics, 
  Department of Physics, Stockholm University, AlbaNova, 
  SE - 106 91 Stockholm --- Sweden}
\email{troms@physto.se}

\author{Julien Lavalle}
\affiliation{Dipartimento di Fisica Teorica,
  Universit\`a di Torino \& INFN,
  via Giuria 1,
  10125 Torino --- Italy}
\email{lavalle@to.infn.it}

\author{Pierre Salati}
\affiliation{LAPTH, Universit\'e de Savoie, CNRS, BP110, F-74941 Annecy-le-Vieux Cedex
 --- France}
\email{salati@lapp.in2p3.fr}

\date{25 May 2009}

\begin{abstract}
\vspace*{-6.4cm}
\mbox{\hspace*{9cm}Preprint: DFTT-23/2008, LAPTH-1313/09}
\vspace*{5.5cm}

The proposal of a galactic population of intermediate mass black holes (IMBHs), 
forming dark matter (DM) ``mini-spikes'' around them, has received 
considerable attention in recent years. In fact, leading in some scenarios to large annihilation 
fluxes in gamma rays, neutrinos and charged cosmic rays, these objects are 
sometimes quoted as one of the most promising targets for indirect DM 
searches. In this letter, we apply a detailed statistical analysis to point 
out that the existing EGRET data already place very stringent limits on those 
scenarios, making it rather unlikely that any of these objects 
will be observed with, e.g., the Fermi/GLAST satellite or upcoming Air 
Cherenkov telescopes. We also demonstrate that prospects for observing signals 
in neutrinos or charged cosmic rays seem even worse. Finally, we address the 
question of whether the excess in the cosmic ray positron/electron flux 
recently reported by PAMELA/ATIC could be due to a nearby DM point 
source like a DM clump or mini-spike; gamma-ray bounds, as well as the recently released Fermi  cosmic ray electron and positron data, again exclude 
such a possibility for conventional DM candidates, and strongly constrain it 
for DM purely annihilating into light leptons.
\end{abstract}

\pacs{95.35.+d, 97.60.Lf, 96.50.S}

\maketitle

%\section{Introduction}
%\label{sec:intro}

The nature of the mysterious DM, vastly dominating the total 
matter content of the universe, still remains unknown.
Particularly plausible candidates, however, are weakly 
interacting massive particles (WIMPs) \cite{review} and indirect DM searches 
aim at discriminating WIMP annihilation products from standard 
astrophysical backgrounds in gamma rays, neutrinos or charged cosmic rays. 

Large DM density enhancements (``mini-spikes'') around IMBHs
 have been proposed as promising targets for indirect DM 
searches in gamma rays \cite{Bertone:2005xz}, where a large number of very 
luminous point sources with identical cutoff in the photon spectrum would 
provide a smoking gun signature. 
Subsequent studies indicated excellent 
observational prospects also for neutrinos \cite{Bertone:2006nq} and charged cosmic rays 
\cite{imbhboost}.
In the following, we restrict ourselves to IMBHs with a mass of around $10^5M_\odot$ that form out of collapsing cold gas in early-forming halos (scenario B of Ref.~\cite{Bertone:2005xz}), 
like in much of the literature on the subject,
since only in this case one arrives at the above mentioned favorable 
prospects for indirect DM detection that have caused considerable recent attention.

While an application of our analysis to other IMBH formation scenarios would be straightforward, it is beyond the scope of this Letter, in which
we critically re-assess the potential of IMBHs for DM 
searches. Our main conclusion is that the most favored DM parameter regions 
are actually already ruled out by the EGRET data \cite{egret3} and that 
configurations predicting a signal in future searches -- and yet being 
consistent with the existing constraints -- are rather unlikely.
For completeness, we will also derive generic limits on any nearby DM point-source.

%\section{A statistical primer}
%\label{sec:stat}

Let us start by recalling some basic statistical properties of an ensemble of 
IMBHs in the galactic halo.
Denoting with $P_c$ the 
probability for a particular IMBH to satisfy some condition $c$, the 
probability for $n$ out of $N$ objects to satisfy $c$ is given by
\be
P_{n,N}(P_c)\equiv \left(\begin{tabular}{c}$N$\\$n$\end{tabular}\right)\,
P_c^n\left(1- P_c\right)^{N-n}\,.
\ee
Since the number of IMBHs in a given realization itself is  a random variable with some distribution 
$p_N$ (in our case a Gaussian with mean $\sim 98$ and variance $\sim 21$ 
\cite{imbhboost}), the probability that $n$ objects in an \emph{arbitrary} 
realization satisfy $c$ becomes
\be
P_n^\mathrm{int}(P_c)\equiv\int_n^\infty {\rm d}N\,p_N(N)\, P_{n,N}(P_c)\,.
\ee

In our context, the relevant astrophysical properties of an IMBH are its 
\emph{distance} $d$ to the Earth and, as a measure of the DM concentration, 
its \emph{annihilation volume}
$
\xi\equiv\int {\rm d}^3x\,\left({\rho(\mathbf{x})}/{\rho_0}\right)^2
$,
where $\rho_0=0.3\,{\rm GeV\,cm^{-3}}$ is the local DM density. In particular, we will be interested in the probability that a single object has $\lambda=\xi/d^2\geq\lambda_0$:
\be
  P_{\lambda\geq\lambda_0}=\int_0^\infty {\rm d}\xi \int_0^{(\xi / \lambda_0)^{1/2}}  \!\!\! \!\!\! {\rm d} d\,
 \big( p_\xi(\xi) p_d(d)\big)\,,
\ee
where we use the (independent) probability densities  $p_d$ and $p_\xi$ to find an IMBH at a distance $d$ from the earth and with a given annihilation volume $\xi$, respectively,  from Ref.~\cite{imbhboost}. These distributions were obtained from Monte Carlo (MC) simulations that follow the evolution of initial mini-spike populations during their orbit in the Milky Way, taking account of possible close encounters. 
Armed with the above notation, we finally arrive at 
\be
  \label{Pfinal}
P^{\rm cum}_n(\lambda_0)=1-\sum_{i=0}^{n-1}P_i^{\rm int}(P_{\lambda\geq\lambda_0})\
\ee
as the probability that a given realization contains \emph{at least} $n$ objects with $\lambda\geq\lambda_0$.

%\section{Gamma ray constraints}
%\label{sec:gamma}
%

The gamma-ray flux from a single IMBH is given by
\be
\label{eq:gamma_flux}
\Phi_{\rm IMBH}^\gamma={N_\gamma\Gamma}/({4\pi d^2})\,,
\ee
where $\Gamma=\frac{1}{2}\sigv \left({\rho_0}/{m_\chi}\right)^2 \xi$ is the 
local injection rate, $m_\chi$ the WIMP mass, $\sigv$ the annihilation 
rate (per unit density) and $N_\gamma$ the number of photons above some energy 
$E_{\rm min}$ per annihilation. 
At $E_\gamma\ll m_\chi$, the photon spectrum looks almost exactly the same for 
generic WIMP DM candidates that annihilate into quark or gauge boson final 
states; if not stated otherwise, we assume annihilation into $b\bar b$ and use {\sc Pythia} \cite{pythia} to compute $N_\gamma$. 
The adiabatic growth of an IMBH redistributes a typical initial DM density distribution 
into a steep profile (a "mini-spike") that is saturated in the innermost region due to 
DM self-annihilations and asymptotically develops an annihilation volume that scales like 
\mbox{$\xi\propto\left(m_\chi/\sigv\right)^{5/7}$} \cite{Bertone:2005xz}. 
Putting in numbers, Eq.(\ref{eq:gamma_flux}) thus becomes
\bea
\label{eq:gamma_flux2}
\Phi_{\rm IMBH}^\gamma&\approx&3.31\cdot10^{-7}
\left(\frac{N_\gamma}{10}\right){\rm cm^{-2}s^{-1}}\\ 
&\times&\left(\frac{\tilde\xi/d^2}{10^5{\rm kpc}}\right)
\bigg(\frac{m_\chi}{{\rm TeV}}\bigg)^{-\frac{9}{7}}
\left(\frac{\sigv }{3\cdot10^{-26}{\rm cm^3s^{-1}}}
\right)^\frac{2}{7}\,, \nonumber
\eea
where $\tilde\xi$ is the annihilation volume for an IMBH when 
$\sigv = 3\cdot10^{-26}{\rm cm^3/s}$ and $m_\chi=1\,{\rm TeV}$.

The upper limit on the flux from point sources not seen by the EGRET satellite strongly 
depends on the position on the sky, but is almost everywhere significantly 
below $\Phi_{\rm max}^\infty\sim2\times10^{-7}\,{\rm cm^{-2}s^{-1}}$ 
for $100\,{\rm MeV}\lesssim E_\gamma\lesssim30\,{\rm GeV}$ \cite{egret3}. On 
the other hand, a considerable number of detected sources with higher fluxes 
remains with no associated low-energy counterpart. Very conservatively assuming 
that all these unidentified sources are, in fact, connected to DM mini-spikes 
would translate into the requirement that at most (1, 4, 10) IMBHs have a flux 
larger than about $\Phi_{\rm max}^{(1,4,10)}\sim(7,\,5.5,\,3.7)\times10^{-7}\,{\rm cm^{-2}s^{-1}}$ 
%($\Phi_{\rm max}^{4}\sim5.5\times10^{-7}\,{\rm cm^{-2}s^{-1}}$, 
%$\Phi_{\rm max}^{1}\sim7\times10^{-7}\,{\rm cm^{-2}s^{-1}}$) 
\cite{pavlidou};
as the spectra of these sources are in almost all cases much softer 
than what is expected from DM annihilation, however, 
this possibility does not appear to be very likely (note also that many of these EGRET sources are not confirmed by the first Fermi data \cite{Abdo:2009mg}).

\begin{figure}[t]
  \includegraphics[width=\columnwidth]{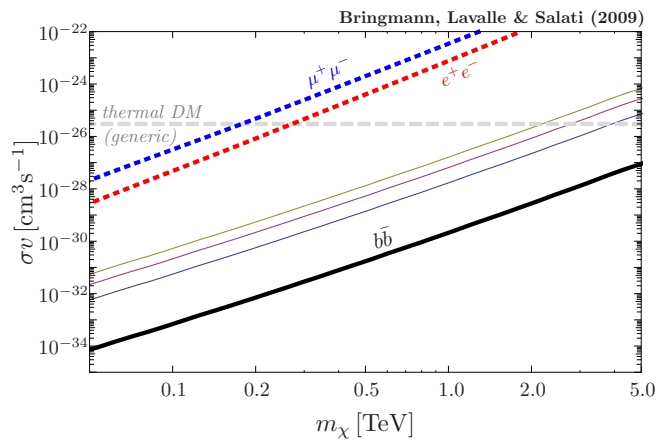}
  \caption{The thin solid lines show the constraints on WIMPs 
    in the IMBH scenario B of \cite{Bertone:2005xz}, from top to bottom deriving from 
    the brightest $10$, $4$ and $1$ object(s), respectively. The region above 
    the thick solid line is excluded by the EGRET limit on unresolved %gamma-ray 
    point 
    sources; the dotted lines show the corresponding 
    constraint if DM were to annihilate only into $e^+e^-$ or $\mu^+\mu^-$. The dashed line 
    indicates the canonical value of 
    $\sigma v=3\cdot10^{-26}\,\mathrm{cm}^3\mathrm{s}^{-1}$ for 
    thermally generated WIMPs. \label{fig_sigmavm}}
\end{figure}

Demanding that at least 5\% of the IMBH realizations should not be in conflict with 
the EGRET constraints on the $n$ brightest sources, we can now use Eq.~(\ref{eq:gamma_flux2}) 
to derive bounds on the annihilation rate and mass of the annihilating 
particles by choosing $\lambda_0=\tilde\xi/d^2$ such that
 $P^{\rm cum}_n(\lambda_0)>0.95$.
 The result is shown in Fig.~\ref{fig_sigmavm} (for similar constraints from the 
 H.E.S.S. experiment, see \cite{hess}). It is interesting to note that the 
$n\sim10$ brightest unidentified EGRET sources lead to similar constraints. 
Having remarked before, however, that an IMBH interpretation of these objects is not too likely,
 we also include for comparison the much tighter 
constraint that results from considering the sensitivity limit of EGRET on 
\emph{unseen} point sources.
We verified these limits in extensive MC simulations and
would like 
to draw special attention to the fact that they lie, indeed, several orders of magnitude 
below the expectation for generic WIMP candidates \cite{review}, 
$m_\chi\sim {\cal O}(100\,{\rm GeV}-1\,{\rm TeV})$ and 
$\sigv \sim {\cal O}(10^{-26}\,{\rm cm^3s^{-1}})$.
Note that the constraints shown in Fig.~\ref{fig_sigmavm} are, in fact even conservative as
the EGRET limit on unresolved point sources is actually in many regions of the sky much smaller than the value used here.

%\section{Maximal neutrino fluxes}
%\label{sec:nu}

The neutrino flux is simply obtained by replacing in 
Eq.~(\ref{eq:gamma_flux}) the index $\gamma$ with $\nu$, so the EGRET limit on gamma rays translates into
\mbox{$\Phi_\mathrm{IMBH}^\nu<(N^\nu \!/ N^\gamma)\Phi_\mathrm{max}^\gamma$}. 
This limit on the
neutrino flux has to be compared with the sensitivity of upcoming km$^3$-sized neutrino 
telescopes \cite{nureview}. Optimistically assuming an effective 
surface area for detection below $1\,$TeV of 
$\sim 0.1 (E_\nu/{\rm TeV})^2 {\rm m^2}$, and 
using the Bartol model for the atmospheric neutrino background~\cite{bartol}, 
we find a background rate of $\sim 10^{-4}$ Hz. A $5\sigma$ detection after 1 
yr would then require a primary neutrino flux 
$\Phi_{\rm IMBH}^\nu \gtrsim 10^{-8}{\rm cm^{-2}s^{-1}}$ (over-optimistically 
assuming  that the target is observed 100\% of the time), which exceeds the 
above stated limit for typical values of $m_\chi$ and $N^\nu$. Of 
course, should DM annihilate mostly into neutrinos,
the EGRET limit would be less 
stringent. However, even in conventional Kaluza-Klein (KK) scenarios 
\cite{ued}, which offer large branching ratios to neutrinos, the EGRET 
constraint still excludes the observation of the neutrino counterpart with 
km$^3$ detectors. Note also that large branching ratios into neutrinos 
actually mean \emph{smaller} $N^\nu$, at least for large $m_\chi$, leading to 
overall worse prospects for detection \cite{Bertone:2006nq}.

%\section{Antimatter cosmic rays}
%\label{sec:crs}

Complementary to gamma rays, antiprotons are a further 
interesting channel of indirect DM detection \cite{pbarDM}. To 
investigate whether the resulting constraints can compete with 
those from gamma rays in the mini-spike scenario, 
we have performed an extensive MC simulation of IMBH realizations. We found that antiprotons are competitive (i.e.~in potential conflict with high-energy $\bar p$ data \cite{PAMpbar}) only for rather heavy WIMPs, with $m_\chi\gtrsim$ 1 TeV, and only when assuming extremely favorable 
propagation parameters, close to the \emph{max} set defined in 
\cite{Donato:2004}; 
outside these somewhat extreme regions of the propagation parameter space, 
large $\bar{p}$ fluxes are generally obtained only in configurations that
 are already excluded by EGRET.

\begin{figure*}[t]
\begin{center}
\includegraphics[width=0.65\columnwidth, clip]{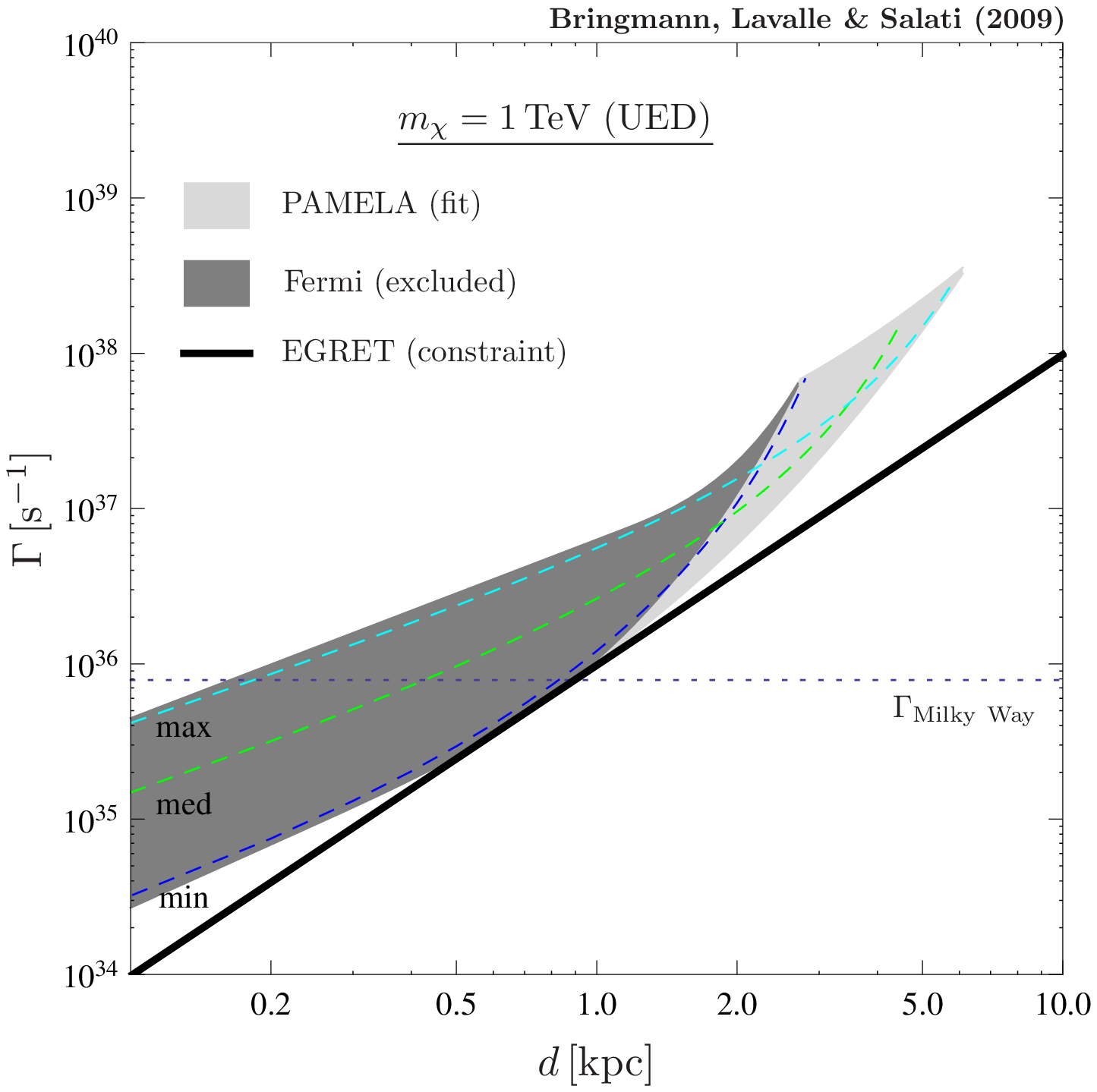}
\includegraphics[width=0.65\columnwidth,clip]{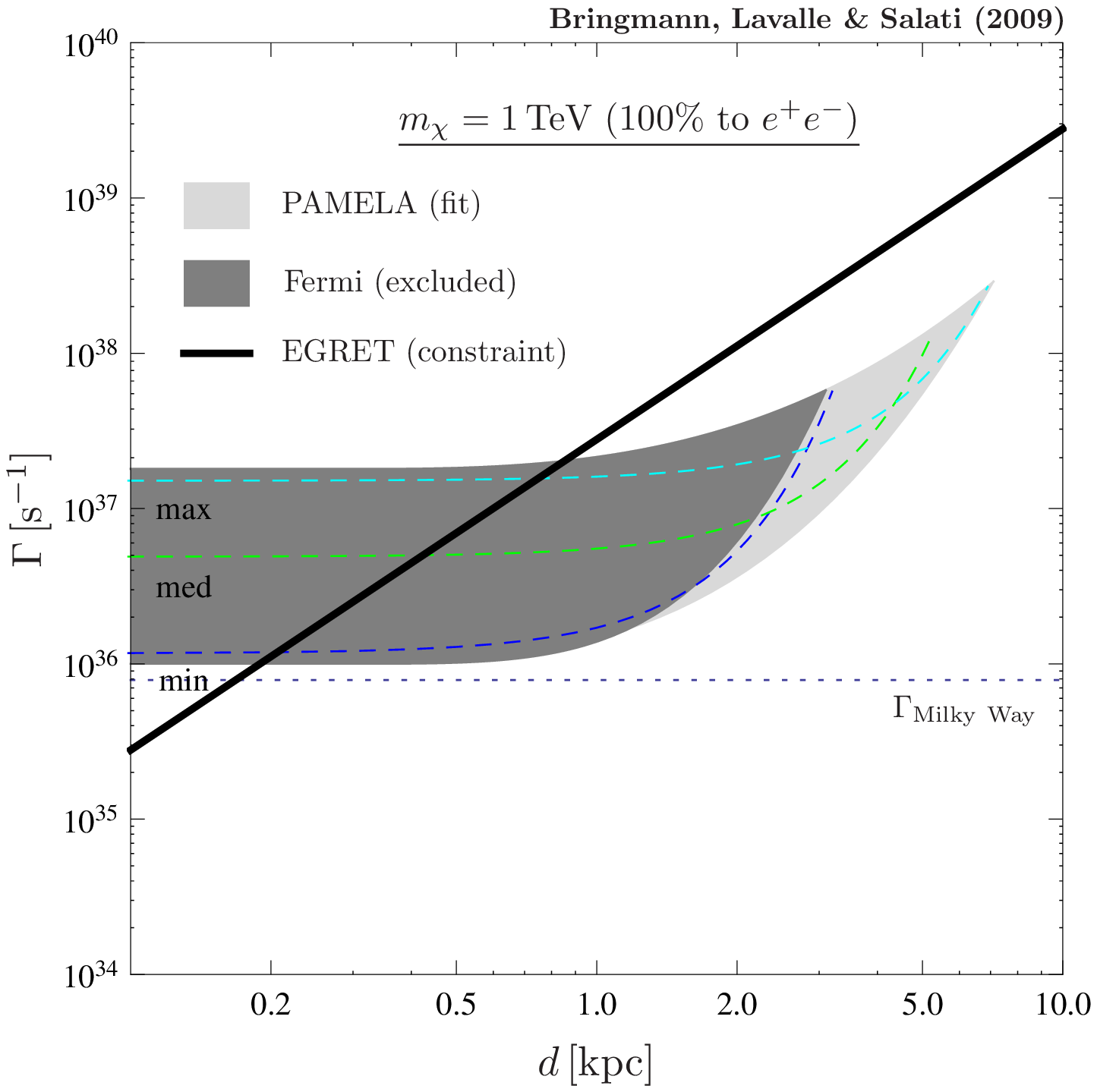}
\includegraphics[width=0.65\columnwidth,clip]{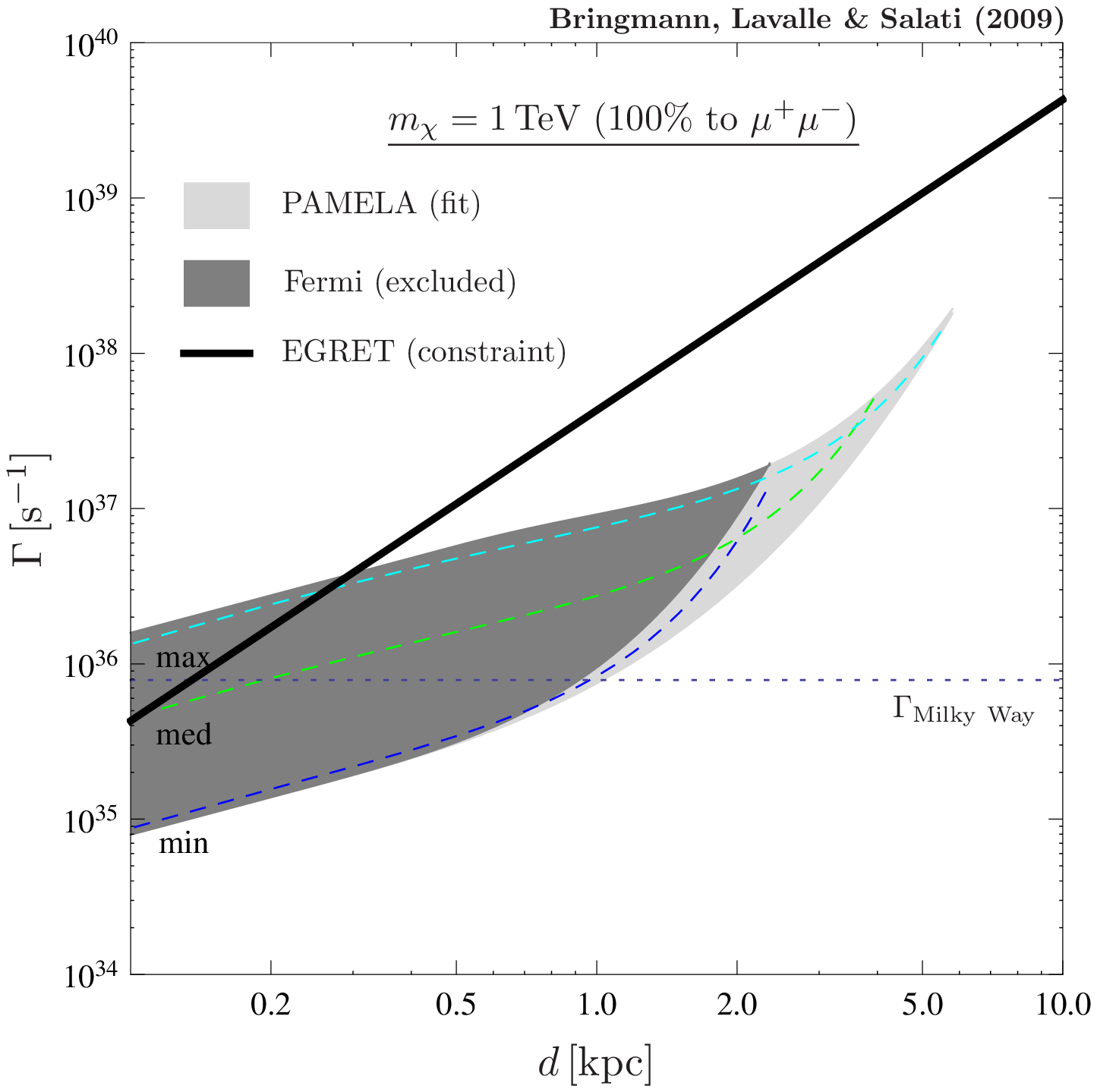}
\caption{\small The solid lines give the EGRET constraints on the DM annihilation 
rate $\Gamma=\frac{1}{2}\sigma v\left({\rho_0}/{m_\chi}\right)^2 \xi$ of a nearby,
generic DM point-source at a distance $d$ from the Earth; from left to right, we 
show the case of KK DM and a fiducial DM candidate annihilating to 
$e^+e^-$ and $\mu^+\mu^-$, respectively. The dashed lines show the $\Gamma$
needed to fit the PAMELA data, for sets of propagation parameters as defined 
in \cite{Delahaye:2008ua}; in the dark shaded area this would produce an $e^\pm$ flux in conflict with the Fermi data at higher energies.
For comparison, the dotted line  indicates 
$\Gamma$ for the \emph{whole Milky Way}, assuming 
$\sigv \sim3\cdot10^{-26}{\rm cm^{-3}s^{-1}}$.}
\label{fig:PAMELA}
\end{center}
\end{figure*}

The excess in cosmic ray positrons and electrons recently reported 
by PAMELA \cite{Adriani:2008zr}, ATIC \cite{atic} and Fermi \cite{Abdo:2009zk}, if interpreted 
in terms of DM annihilation, points at DM masses in the TeV range with an unusually large
branching ratio into light leptons \cite{Cirelli:2008pk}. In 
order to fit the data, however, standard DM candidates need extremely large 
boost factors that are not expected in current models of structure formation 
\cite{Springel:2008cc},  corresponding to effective annihilation rates $\sim10^3$ times the 
generic value for thermally produced DM.  
For comparison, we thus furthermore include in Fig.~\ref{fig_sigmavm} the extreme 
situation of DM particles annihilating only into light leptons,
% (while certainly not 
%standard, such DM candidates have recently been proposed in, e.g., 
%\cite{aniHamed:2008qn}),
 in which case much less 
photons are produced and we can 
apply the analytic expression for $N_\gamma$ given in \cite{UEDgamma};
even in this somewhat contrived situation low-mass models turn out to be difficult to realize,
 at least for strongly enhanced annihilation rates,
making it very unlikely for, e.g, Fermi-LAT \cite{glast} to see the expected cutoff in the photon spectrum for an IMBH that 
has escaped detection by EGRET.

Let us now turn to the
possibility that the large required annihilation flux can be attributed to a nearby high DM 
concentration, like for example a DM clump or a DM mini-spike around an IMBH, 
located at some  distance $d$ to the Earth.  In Fig.~\ref{fig:PAMELA}, 
we use Eq.~(\ref{eq:gamma_flux}) and the EGRET sensitivity limit
to constrain the annihilation rate $\Gamma$ of such a generic DM point source. 
In the same figure, we indicate as a gray area in the $\Gamma-d$ plane the annihilation rate that would be necessary to 
explain the PAMELA data, taking into account the allowed range for the 
$e^+/e^-$ propagation parameters (consistent with both the background and the 
signal) by using the \emph{min/med/max} configurations of 
\cite{Delahaye:2008ua}. The dark shaded region, finally,  shows the rather conservative constraint on these combinations of $\Gamma$ and $d$ that arises from requiring that the positron and electron flux \emph{from DM annihilation alone} should not exceed the Fermi data.
For comparison, we separately consider the case of KK DM,
as an example of a standard WIMP with exceptionally large branching ratios
into light leptons,
as well as DM only annihilating to $e^+e^-$ or $\mu^+\mu^-$.

The first important observation is that standard WIMPs, which usually feature
smaller  branching ratios into light 
leptons than KK DM, basically cannot account for the PAMELA/ATIC data in this way without 
violating the EGRET bounds -- the reason being the still relatively large 
contribution from non-leptonic channels to the photon spectrum at low energies.
However, if one takes a more phenomenological approach and allows 
DM particles annihilating at 100\% into $e^+e^-$ or $\mu^+\mu^-$ pairs, 
one would technically be able to fit the  data by placing a DM
point source at a distance $2\,\mathrm{kpc}\lesssim d\lesssim5\,\mathrm{kpc}$ --
at the price of requiring an enormously bright object,
\emph{more luminous than the whole Milky Way}! 
This would correspond to a DM clump of mass $\gtrsim 10^{11}M_\odot$ in 
conventional cosmological scenarios \cite{Lavalle:2008}. 
Since finding such a massive clump relatively close to the Earth is extremely 
unlikely \cite{Springel:2008cc}, we arrive at a considerably more pessimistic
conclusion than recently obtained by \cite{Hooper:2008} about the possibility
of explaining the PAMELA data in terms of a nearby clump of annihilating WIMPs.
Mini-spikes, on the other hand, \emph{are} extremely bright: for  DM
with $\sigv\sim3\cdot10^{-26}{\rm cm^{3}s^{-1}}$ in the IMBH scenario discussed
earlier, e.g., the probability to encounter 
at least one object inside the light gray area of the 
middle (right) panel of Fig.~\ref{fig:PAMELA} is roughly 84\% (37\%). 
For 
a -- certainly non-standard -- WIMP candidate annihilating almost exclusively 
into $e^+e^-$ or $\mu^+\mu^-$, a galactic population of IMBHs might thus 
indeed provide a positron flux large enough to fit the data and yet be consistent
with present-day constraints.  Note, however, that this conclusion does \emph{not} hold for DM candidates with \emph{intrinsically} enhanced annihilation rates: already for $\sigv\sim10^{-24}{\rm cm^{3}s^{-1}}$, the above quoted probabilities drop to 0.9\% (0.1\%).
% for $e^\pm$ ($\mu^\pm$) final states.

%\section{Conclusions}

In this Letter, we have reconsidered the prospects for indirect DM detection 
in the presence of a galactic population of IMBHs and found them not to be 
very promising given that existing data, in particular from gamma rays, already place severe constraints on the scenario.
While it was noted before that already EGRET should have seen 
some of these objects \cite{Bertone:2005xz}, the resulting constraints were 
not taken into account in the subsequent studies of, e.g., 
\cite{Bertone:2006nq,imbhboost}. This is the main reason for the discrepancy 
between our pessimistic and earlier rather optimistic conclusions about 
possible effects of DM mini-spikes on indirect DM searches. While beyond the scope of this letter,
it would be interesting to extend the study presented here and apply a consistent treatment 
of all available constraints also to other IMBH formation scenarios in order to predict realistic prospects for indirect DM detection.

We have also addressed the possibility of explaining the recent PAMELA 
observations in terms of DM annihilation by placing a dark object in close 
vicinity to the Earth. While this option is ruled out from the gamma-ray 
constraints for standard WIMP candidates, a nearby IMBH (but not an ordinary 
DM clump) may in principle provide a sufficient amount of positrons if one 
assumes that the DM particles annihilate purely into light leptons. 
Even this seemingly a bit far-fetched scenario could soon be ruled out if 
Fermi \cite{glast} does not observe any corresponding point-sources in 
gamma rays. 
We would like to use this opportunity to recall that a very plausible hypothesis
for  the PAMELA results is anyway a nearby 
pulsar, i.e.~an explanation in terms of astrophysics rather than DM 
 annihilation \cite{PAMastro}.

%\section*{Acknowledgements}
\paragraph{Acknowledgements:}
We greatly thank Gianfranco Bertone for letting us use his catalogue of 
IMBH sources and the anonymous referees for very useful suggestions. TB and JL 
thank LAPTH for hospitality when this work was 
initiated. This work is supported, in part, by the French ANR project 
ToolsDMColl (BLAN07-2-194882). PS thanks the Perimeter Institute for
hospitality when this work was finished.

%%%%%%%%%%%%%%%%%%%%%%%%%%%%%%%%%%%%%%%%%%%%%%%%%%%%%%%%%%%%%%%


\begin{thebibliography}{00}

\bibitem{review}
 G.~Jungman, M.~Kamionkowski and K.~Griest,
 %``Supersymmetric d matter,''
 Phys.\ Rept.\  {\bf 267}, 195 (1996);
% [hep-ph/9506380];
 %%CITATION = PRPLC,267,195;%%
 L.~Bergstr\"om,
 %``Non-baryonic dark matter: Observational evidence and detection methods,''
 Rept.\ Prog.\ Phys.\  {\bf 63}, 793 (2000);
% [hep-ph/0002126];
  G.~Bertone, D.~Hooper and J.~Silk,
  %``Particle dark matter: Evidence, candidates and constraints,''
  Phys.\ Rept.\  {\bf 405}, 279 (2005).
%  [hep-ph/0404175].
  %%CITATION = HEP-PH 0404175;%%


\bibitem{Bertone:2005xz}
  G.~Bertone, A.~R.~Zentner and J.~Silk,
  %``A new signature of dark matter annihilations: Gamma-rays from
  %intermediate-mass black holes,''
  Phys.\ Rev.\  D {\bf 72}, 103517 (2005).
%  [astro-ph/0509565];
  %%CITATION = PHRVA,D72,103517;%%
  M.~Fornasa and G.~Bertone,
  %``Black Holes as Dark Matter Annihilation Boosters,''
  Int.\ J.\ Mod.\ Phys.\  D {\bf 17}, 1125 (2008).
%  [arXiv:0711.3148 [astro-ph]].
  %%CITATION = IMPAE,D17,1125;%%

\bibitem{Bertone:2006nq}
  G.~Bertone,
  %``Prospects for detecting dark matter with neutrino telescopes in
  %intermediate mass black holes scenarios,''
  Phys.\ Rev.\  D {\bf 73}, 103519 (2006).
%  [astro-ph/0603148].
  %%CITATION = PHRVA,D73,103519;%%

\bibitem{imbhboost}
  P.~Brun \emph{et al.},
%  G.~Bertone, J.~Lavalle, P.~Salati and R.~Taillet,
  %``Antiproton and Positron Signal Enhancement in Dark Matter Mini-Spikes
  %Scenarios,''
  Phys.\ Rev.\  D {\bf 76} (2007) 083506.
%  [arXiv:0704.2543 [astro-ph]].
  %%CITATION = PHRVA,D76,083506;%%

\bibitem{egret3}
  Hartman et al., \emph{Ap.J.S.} {\bf 123}, 79 (1999).

\bibitem{pythia}
  T.~Sj\"ostrand, S.~Mrenna and P.~Skands,
  %``PYTHIA 6.4 physics and manual,''
  JHEP {\bf 0605}, 026 (2006)
  [arXiv:hep-ph/0603175].
  %%CITATION = JHEPA,0605,026;%%

\bibitem{pavlidou}
  V.~Pavlidou \emph{et al.},
%  J.~M.~Siegal-Gaskins, C.~Brown, B.~D.~Fields and A.~V.~Olinto,
  %``Unidentified EGRET Sources and the Extragalactic Gamma-Ray Background,''
  Astrophys.\ Space Sci.\  {\bf 309} (2007) 81.
%  [astro-ph/0611271].
  %%CITATION = APSSB,309,81;%%

\bibitem{Abdo:2009mg}
  A.~A.~Abdo {\it et al.}
  %  [Fermi LAT Collaboration],
  %``Fermi Large Area Telescope Bright Gamma-ray Source List,''
  arXiv:0902.1340 [astro-ph.HE].
  %%CITATION = ARXIV:0902.1340;%%

\bibitem{hess}
  F.~Aharonian {\it et al.},
%  [HESS Collaboration],
  %``Search for Gamma-rays from Dark Matter annihilations around Intermediate
  %Mass Black Holes with the H.E.S.S. experiment,''
  Phys.\ Rev.\  D {\bf 78}, 072008 (2008).
%  [arXiv:0806.2981 [astro-ph]].
  %%CITATION = PHRVA,D78,072008;%%

\bibitem{nureview}
  See, e.g., 
  T.~Montaruli,
  %``Review on Neutrino Telescopes,''
  arXiv:0901.2661 [astro-ph].
  %%CITATION = ARXIV:0901.2661;%%

\bibitem{bartol}
  G.~Barr, T.~K.~Gaisser and T.~Stanev, 
  Phys. Rev. D {\bf 39} (1989).

\bibitem{ued}
  D.~Hooper and S.~Profumo,
  %``Dark matter and collider phenomenology of universal extra dimensions,''
  Phys.\ Rept.\  {\bf 453}, 29 (2007).
%  [hep-ph/0701197].
  %%CITATION = PRPLC,453,29;%%

\bibitem{pbarDM}
  T.~Bringmann and P.~Salati,
  %``The galactic antiproton spectrum at high energies: background   expectation
  %vs. exotic contributions,''
  Phys.\ Rev.\  D {\bf 75}, 083006 (2007);
%  [astro-ph/0612514];
  %%CITATION = PHRVA,D75,083006;%%
  F.~Donato {\it et al.},
%  D.~Maurin, P.~Brun, T.~Delahaye and P.~Salati,
  %``Constraints on WIMP Dark Matter from the High Energy PAMELA $\bar{p}/p$
  %data,''
  Phys.\ Rev.\ Lett.\  {\bf 102}, 071301 (2009).
%  [arXiv:0810.5292 [astro-ph]].
  %%CITATION = PRLTA,102,071301;%%
  
\bibitem{PAMpbar}
 O.~Adriani {\it et al.},
  %``A new measurement of the antiproton-to-proton flux ratio up to 100 GeV in
  %the cosmic radiation,''
  Phys.\ Rev.\ Lett.\  {\bf 102}, 051101 (2009).
%  [arXiv:0810.4994 [astro-ph]].
  %%CITATION = PRLTA,102,051101;%%

\bibitem{Donato:2004}
  F.~Donato {\it et al.},
%  N.~Fornengo, D.~Maurin, P.~Salati and R.~Taillet,
  Phys. Rev. D {\bf 69}, 06351 (2004).
%  [astro-ph/0306207].

\bibitem{Adriani:2008zr}
  O.~Adriani {\it et al.},
  %``Observation of an anomalous positron abundance in the cosmic radiation,''
  arXiv:0810.4995 [astro-ph].
  %%CITATION = ARXIV:0810.4995;%%

\bibitem{atic}
  J.~Chang {\it et al.}, Nature {\bf 456}, 362 (2008).

\bibitem{Abdo:2009zk}
  A.~A.~Abdo {\it et al.}
  %  [The Fermi LAT Collaboration],
  %``Measurement of the Cosmic Ray e+ plus e- spectrum from 20 GeV to 1 TeV with
  %the Fermi Large Area Telescope,''
  Phys.\ Rev.\ Lett.\  {\bf 102}, 181101 (2009).
%  [arXiv:0905.0025 [astro-ph.HE]].
  %%CITATION = PRLTA,102,181101;%%

\bibitem{Cirelli:2008pk}
  see, e.g., M.~Cirelli, 
% {\it et al.},
  M.~Kadastik, M.~Raidal and A.~Strumia,
  %``Model-independent implications of the e+, e-, anti-proton cosmic ray
  %spectra on properties of Dark Matter,''
  Nucl.\ Phys.\  B {\bf 813}, 1 (2009).
%  [arXiv:0809.2409 [hep-ph]].
  %%CITATION = NUPHA,B813,1;%%


\bibitem{Springel:2008cc}
  V.~Springel {\it et al.},
  %``The Aquarius Project: the subhalos of galactic halos,''
  MNRAS {\bf 391}, 1685 (2008).
%  [arXiv:0809.0898 [astro-ph]].
  %%CITATION = ARXIV:0809.0898;%%

%\bibitem{ArkaniHamed:2008qn}
%  N.~Arkani-Hamed, 
%  {\it et al.},
%  D.~P.~Finkbeiner, T.~Slatyer and N.~Weiner,
  %``A Theory of Dark Matter,''
%  Phys. Rev. D {\bf 79}, 015014 (2009).
%  [arXiv:0810.0713 [hep-ph]].
  %%CITATION = ARXIV:0810.0713;%%

\bibitem{UEDgamma}
  L.~Bergstr\"om,
% {\it et al.},
  T.~Bringmann, M.~Eriksson and M.~Gustafsson,
  %``Gamma rays from Kaluza-Klein dark matter,''
  Phys.\ Rev.\ Lett.\  {\bf 94}, 131301 (2005).
%  [astro-ph/0410359].

\bibitem{glast}
  E.~A.~Baltz {\it et al.},
  %``Pre-launch estimates for GLAST sensitivity to Dark Matter annihilation
  %signals,''
  JCAP {\bf 0807}, 013 (2008).
%  [0806.2911 [astro-ph]].
  %%CITATION = JCAPA,0807,013;%%

\bibitem{Delahaye:2008ua}
  T.~Delahaye {\it et al.},
%  F.~Donato, N.~Fornengo, J.~Lavalle, R.~Lineros, P.~Salati and R.~Taillet,
  %``Galactic secondary positron flux at the Earth,''
  arXiv:0809.5268 [astro-ph].
  %%CITATION = ARXIV:0809.5268;%%

\bibitem{Lavalle:2008}
  J.~Lavalle, Q.~Yuan, D.~Maurin and X.~J.~Bi,
  Astron. Astrophys. {\bf 479}, 427 (2008).
%  arXiv:0709.3634 [astro-ph].

\bibitem{Hooper:2008}
  D.~Hooper, A. Stebbins and K.~M.~Zurek,
  arXiv:0812.3202 [astro-ph].

\bibitem{PAMastro}
  A.~Boulares, ApJ {\bf 342}, 807 (1989);
  F.~A.~Aharonian, A.~M.~Atoyan and H.~J.~Voelk, 
  Astron. Astrophys. {\bf 294}, L41-L44 (1995);
  H.~Y\"uksel, M.~D.~Kistler and T.~Stanev, arXiv:0810.2784;
  D.~Hooper, P.~Blasi and P.~D.~Serpico,
  %``Pulsars as the Sources of High Energy Cosmic Ray Positrons,''
  JCAP {\bf 0901}, 025 (2009);
%  [arXiv:0810.1527 [astro-ph]];
  %%CITATION = JCAPA,0901,025;%%  
  S.~Profumo,
  %``Dissecting Pamela (and ATIC) with Occam's Razor: existing, well-known
  %Pulsars naturally account for the 'anomalous' Cosmic-Ray Electron and
  %Positron Data,''
%  arXiv:0812.4457;
  %%CITATION = ARXIV:0812.4457;%%
  N.~J.~Shaviv, E.~Nakar and T.~Piran,
  %``Natural explanation for the anomalous positron to electron ratio with
  %supernova remnants as the sole cosmic ray source,''
  arXiv:0902.0376 [astro-ph.HE].
  %%CITATION = ARXIV:0902.0376;%%
\end{thebibliography}
\end{document}